\documentstyle[12pt]{article}
\advance\hoffset by -1.1truecm
\advance\voffset by -1.0truecm

\newcommand{\gapproxeq}{\lower .7ex\hbox{$\;\stackrel{\textstyle >}{\sim}\;$}}
\newcommand{\lapproxeq}{\lower .7ex\hbox{$\;\stackrel{\textstyle <}{\sim}\;$}}
\newcommand{\eqn}[1]{(\ref{#1})}
\newcommand{\be}{\begin{equation}}
\newcommand{\ee}{\end{equation}}

\newcommand{\bea}{\begin{eqnarray}}
\newcommand{\eea}{\end{eqnarray}}
\newcommand{\bean}{\begin{eqnarray*}}
\newcommand{\eean}{\end{eqnarray*}}

\newcommand{\I}{\mbox{\rm I} \hspace{-0.5em} \mbox{\rm I}\,}

\newcommand{\complex}{
        \mbox{C \hspace{-1.16em} \raisebox{-0.018em}{\sf l}}\;}

\newcommand{\quater}{\mbox{I \hspace{-0.86em} H}}

\newcommand{\alg}{\mbox{${\cal A}$}}

\begin{document}
\hfill DSF 45/96, OUTP-96-61-P, hep-th/9610035\\

\begin{center}

{\Large \bf Fermion Hilbert Space and Fermion Doubling in the Noncommutative
Geometry Approach to Gauge Theories}

\vskip 2cm

{\bf \large F. Lizzi${}^{1,2}$,
G. Mangano{}$^{1}$,
G. Miele${}^{1,2}$ and
G. Sparano${}^{1}$}

\vskip 1.5cm

${}^1$ {\it  Dipartimento di Scienze Fisiche, Universit\`a di Napoli {\sl Federico II},\\
and INFN, Sezione di Napoli, Italy}\\
\ \\
and\\
\ \\
${}^2$ {\it Theoretical Physics, 1 Keble Road\\ Oxford, OX1 3NP, UK}

\end{center}

\begin{abstract}
In this paper we study the structure of the Hilbert space for the
recent noncommutative geometry models of gauge theories. We point out
the presence of unphysical degrees of freedom similar to the ones
appearing in lattice gauge theories (fermion doubling). We investigate
the possibility of projecting out these states at the various levels
in the construction, but we find that the results of these attempts
are either physically unacceptable or geometrically unappealing.
\end{abstract}

\vfill
\noindent
Lizzi@na.infn.it\\
Mangano@na.infn.it\\
Miele@na.infn.it\\
Sparano@na.infn.it
\newpage

\section{Introduction}
Noncommutative geometry \cite{Book} provides a powerful algebraic scheme
to handle a large variety of geometrical frameworks.  Its application to
gauge theories, and in particular to the Standard Model (SM) of strong and
electroweak forces \cite{Book}--\cite{Mars}, is a unique original
way to fully
geometrize the interaction of elementary particles. More recently,
attempts have been made towards an unification with gravity as well
\cite{Connesgrav,ChamsConnes}. In noncommutative geometry the role that
classically is played by a manifold, seen as an ensemble of points, is
taken by a $*$-algebra, which in the {\it commutative} case is just the
algebra of continuous complex valued functions, but in general can be a
generic non abelian $*$-algebra. This algebra is then represented as
bounded operators on an Hilbert space on which a generalized Dirac
operator $D$ also acts, providing all information usually carried by a
metric structure \cite{Book}.

A very appealing aspect of the Connes--Lott (CL) version of the SM
and of his subsequent versions and improvements (for a review see
\cite{Cordelia}) is
that the Hilbert space on which the algebra and the generalized Dirac
operators act is the space of physical fermions. In the model the
fermionic action corresponds to a generalization of the (interaction)
Dirac action $\overline{\psi} ( \partial\!\!\!/ + A\!\!\!/ ) \psi$ while
the bosonic one is obtained by taking the trace of the squared curvature
two-form which is constructed out of the algebra
$C^{\infty} (M, \complex)\otimes \alg_F$, where $C^{\infty}(M,\complex)$
is the algebra of smooth
complex valued function on the (euclidean) 4-dimensional space--time
manifold $M$, and
$\alg_F = M_3 \oplus \quater \oplus \complex$, with $M_3$
and $\quater$ the algebra of 3$\times$3 complex matrices and of
quaternions, respectively.
The result generalizes the Yang--Mills
euclidean action $(1/4) F^{\mu\nu}F_{\mu\nu}$. Most remarkably, in this
framework, the scalar Higgs field appears as the connection in the
internal (noncommutative) space while its action, including the usually
{\it ad hoc} quartic potential, naturally appears as the square of the 
curvature in the internal space. This model is quite constrained and, with 
the choice of the Hilbert space composed by the known fermions only, seems to
point to some unique features of the SM, forbidding for example
standard grand unified theories \cite{GUT}.

Quite recently Chamseddine and Connes (CC) \cite{ChamsConnes} have also
proposed a different definition of the bosonic action, based on the
so-called {\it spectral action principle}, which from the generalized
Dirac operator only, now including also the gravitational spin
connections, produces the SM action coupled to Einstein plus
Weyl gravity.

In both approaches the Hilbert space of fermions ${\cal H}$ seems to play
a crucial role. On one side it is necessary in order to represent the
algebra, which gives the topology of space, though this last feature can
be recovered independently of whether an explicit representation is
assigned or not. On the other side ${\cal H}$ is definitely necessary for
the introduction of the $D$ operator, which encodes, as mentioned,
all information on the metric.

The structure chosen for ${\cal H}$ is the one of a tensor product of a
continuous infinite dimensional factor, the space $L^2(S_M)$ of square
integrable Dirac spinor over $M$, which is related to space--time, times a
finite dimensional space, which describes the physical particle degrees of
freedom, {\it including helicity}. In this paper we point out that, for a
chiral gauge theory built up in both the CL and CC approaches, this choice
for ${\cal H}$, though imposed by the initial ansatz for the algebra,
introduces a problem of overcounting of the physical degrees of
freedom. 
This is basically due to the fact that
the helicity degrees of freedom are contained both in the spinor,
and in the finite dimensional space. More seriously, in chiral gauge
theory, some of these degrees of freedom are unphysical. They look like
mirror fermions, namely states which couple to the chiral factors of the
gauge group but which have, however, wrong chiral quantum numbers.
This is in a way analogous to the doubling of degrees of freedom
encountered in lattice gauge theories.
For the
case of the minimal SM, for example, we will see how
right-handed Dirac spinors would be coupled in general to the $SU(2)_L$
gauge bosons.

Some of the spurious states are a consequence of the {\it duplication} of
degrees of freedom in the finite part of the Hilbert space. In all gauge
models, in fact, in which more than one factor of the group acts on some
fermion multiplet, in order to correctly represent the algebra, it is
necessary to include for all fermions the corresponding charge--conjugated
states. Once the action is obtained a simple {\it
identification} procedure, which is to say, a quotient of the initial
Hilbert space, is sufficient to remove this unwanted duplication of
particles\footnote{It is worth noticing at this point that such a problem 
is absent in the so called `old version' of the model \cite{Book,JoePepe}.}.

On the other side the presence of mirror fermions, as we will see in the
following, seems more closely related to the choice for ${\cal H}$ in the
form of a tensor product. What is implicitly done in the literature is
to throw away at the very end the unwanted terms from the action. One wonders
therefore if there is a more geometrical way of getting rid of the
unphysical fermions. Actually they can be eliminated
by a {\it projection} onto a proper subspace of ${\cal H}$. This
projection, however, may lead to different results if performed at
different stages of the noncommutative geometric construction of a gauge
theory, since the corresponding projector operators do not commute with
all the intermediate steps of such a construction. For example, since they
project onto definite chirality states in $L^2(S_M)$, they do not commute
with the Dirac operator, which contains Dirac gamma matrices, as well as
with the connection one-forms.

As we will see in the following, using the projection at the level of the 
algebra leads to a trivial result.
It is also possible to perform the projection, for the bosonic part, at
the level of the curvature two-form. In this way the action
is obtained by tracing the squared curvature over the physical Hilbert
subspace only. As far as the fermionic action is concerned, this 
procedure gives the same result obtained in the literature
Instead, in the bosonic sector, in addition to the usual kinetic
term, topological terms of the form $\epsilon_{\mu \nu
\rho \sigma} F^{\mu \nu} F^{\rho \sigma}$ 
will generally appear. However,
the combination of the two terms in the curvature is such that only the
self--dual or anti self--dual components of the field survive. This means
that half of the gauge physical degrees of freedom has been projected out
as well.

It seems therefore that the only consistent procedure to obtain the action
of the SM and, more generally, of any chiral gauge theory,
is to just
neglect the unwanted mirror states in the action. This ad hoc restriction
of the action just at the very end of the powerful noncommutative
construction is quite unsatisfactory and probably, a different, less
trivial choice for the structure of the fermion Hilbert space is required,
possibly on the lines of a supersymmetric generalization \cite{KalauWalze} 
or some even more radical changes \cite{nonassociative}.
 
The paper is organised as follows: in section 2 we review the CL
construction and the concept of spectral triple. We also discuss the problem
of redundant fermionic degrees of freedom and, in particular, of
mirror fermions. In section 3 we first discuss a simple model based onto
a spontaneously broken $SU(2)_L \otimes SU(2)_R$ gauge symmetry, showing how
the possible projections work.
The case of the Standard Model is then considered.
Section 4 is devoted to a similar study in the
new CC model for the simple $SU(2)_L \otimes SU(2)_R$ unbroken case.
Finally, in section 5 we give our conclusions and outlook.

\section{Spectral Triple, the Construction of Gauge Theories and
Fermion Hilbert Space} 
The basic ingredient of the noncommutative Geometry construction is the
so-called {\it spectral triple}, denoted by $({\cal A},{\cal H},D)$, where
${\cal A}$ is an involutive $*$-algebra faithfully represented by bounded
operators on the Hilbert space ${\cal H}$, and $D$ is a selfadjoint
operator with compact resolvent (generalized Dirac operator). The spectral
triple becomes a {\it real spectral triple} if an antilinear isometry $J$
of ${\cal H}$, obeying suitable relations is introduced \cite{RealNCG}.
Note that $J$ can be seen as a generalized {\sl CPT} operator.

In this framework a gauge theory, with group of invariance $G$, is fully
geometrized and it is on the same footing as gravity. The former, in
fact, emerges as the gauge theory of the inner automorphisms of the
algebra 
\begin{equation}
{\cal A}=C^{\infty}({ M}, \complex) \otimes {\cal A}_{F}~~~,\label{2.1}
\end{equation}
where ${\cal A}_{F}$ is the smallest $*$-algebra containing $G$ as the
group of its unitary elements. Analogously, the latter can be seen as
the gauge theory of diffeomorphisms of ${ M}$, which are nothing
but the outer automorphisms of ${\cal A}$. 

As far as the Hilbert space ${\cal H}$ is concerned a suitable choice
is to take 
\be
{\cal H} = L^2(S_M) \otimes {\cal H}_F~~~,
\label{2.2}
\ee
where $L^2(S_M)$ is the space of square integrable spinors defined on
$M$, and ${\cal H}_F$ is a finite dimensional linear space
corresponding to all discrete degrees of freedom, like chirality,
flavour, charge etc. Finally, the generalized Dirac operator is 
\be
D=\partial\!\!\!/ \ \otimes \I + \gamma_5 \otimes D_F~~~,
\label{2.3}
\ee
with $D_F$ denoting the selfadjoint fermion mass matrix. 
In this way, the real
spectral triples $({\cal A},{\cal H},D)$ is the tensor product of two
real triples, one which is the continuous (space-time) part,
$(C^{\infty}({ M,\complex}),L^2(S_M),\partial\!\!\!/)$ and another for
the internal part,
$({\cal A}_{F}, {\cal H}_F, D_F)$. 

Given a generalized Dirac operator $D$, the gauge
connection is written as 
\begin{equation}
A = \sum_{i} \beta_i [D,\alpha_i] \equiv \sum_{i} \beta_i d \alpha_i~~~,
\label{2.4}
\end{equation}
where $\alpha_i$ and $\beta_i$ are elements of ${\cal A}$ such that
$A$ is hermitian, and the differential $d$ is defined by $d \alpha
\equiv [D,\alpha]$. From the connection $A$, one defines the
curvature $\theta$ as\footnote{Note that the $d$ operator so defined
is not nilpotent and hence a quotient is necessary in order to 
obtain the correct differential algebras \cite{Book,JoePepe}.
The forms quotiented out are the so called junk forms.}
\begin{equation}
\theta \equiv dA + A^2~~~,\label{2.5}
\end{equation}
and thus the bosonic action is obtained as 
\begin{equation}
{\cal S_{B}} =  \mbox{Tr}\,\theta^2~~~.\label{2.6}
\end{equation}
Note that the trace includes the integration over $M$.
For the fermionic action one has instead
\begin{equation}
{\cal S_{F}} = \langle\psi,(D + A + JAJ)\psi\rangle~~~.\label{2.6a}
\end{equation}
It is worth noticing that exterior algebra emerges from the Clifford
algebra as its antisymmetric part, and thus it is crucial that the
Hilbert space contains as its continuous part the space of Dirac
spinors $L^2(S_M)$. In fact, the use of Weyl spinors on a four
dimensional space-time $M$ would lead to an incorrect result, since the
algebra generated by the Pauli matrices plus identity is only four
dimensional, and thus is not sufficient to faithfully represent the
corresponding 15-dimensional exterior algebra. In particular, all the
two forms would be junk forms and thus $\theta$ would be trivial. 

The choice (\ref{2.2}) for the Hilbert space has
problems in case of theories, like the Standard Model, where
fermions with different chirality transform independently under the
gauge group. In the new formulation of the SM \'a la Connes and Lott
the discrete part of the Hilbert space ${\cal H}_F$ results 
\be
{\cal H}_F = {\cal H}_L\oplus {\cal H}_R\oplus {\cal H}^c_R\oplus 
{\cal H}^c_L~~~, 
\label{2.7}
\ee
where
\bea
{\cal H}_L & = & 
\left(\complex^2 \otimes \complex^N \otimes \complex^3 \right)
\oplus \left(\complex^2 \otimes \complex^N \otimes \complex \right) ~~~,
\label{2.8}\\
{\cal H}_R & = & 
\left(\left(\complex \oplus \complex\right) \otimes \complex^N 
\otimes \complex^3 \right)
\oplus \left(\complex \otimes \complex^N \otimes \complex \right) ~~~,
\label{2.9}
\eea
and ${\cal H}^c_{L,R}$ are the corresponding spaces for antiparticles
\bea
{\cal H}^c_R & = & 
\left(\complex^2 \otimes \complex^N \otimes \complex^3 \right)
\oplus \left(\complex^2 \otimes \complex^N \otimes \complex \right) ~~~,
\label{2.10}\\
{\cal H}^c_L & = & 
\left(\left(\complex \oplus \complex\right) \otimes \complex^N 
\otimes \complex^3 \right)
\oplus \left(\complex \otimes \complex^N \otimes \complex \right) ~~~.
\label{2.11}
\eea
In this framework a natural basis is given by
\bea & 
\left(\begin{array}{c} u_{\alpha} \\ d_{\alpha} \end{array}\right)_L~,~
\left(\begin{array}{c} c_{\alpha} \\ s_{\alpha} \end{array}\right)_L~,~
\left(\begin{array}{c} t_{\alpha} \\ b_{\alpha} \end{array}\right)_L~,~
\left(\begin{array}{c} \nu_{e} \\ e \end{array}\right)_L~,~
\left(\begin{array}{c} \nu_{\mu} \\ \mu \end{array}\right)_L~,~
\left(\begin{array}{c} \nu_{\tau} \\ \tau \end{array}\right)_L~,~
\label{2.12}\\ 
& & \nonumber\\
& \begin{array}{c} (u_{\alpha})_R, \\ (d_{\alpha})_R, \end{array}~
\begin{array}{c} (c_{\alpha})_R, \\ (s_{\alpha})_R, \end{array}~
\begin{array}{c} (t_{\alpha})_R \\  (b_{\alpha})_R, \end{array}~
(e)_R,~(\mu)_R,~(\tau)_R,
\label{2.13}\\ 
& & \nonumber\\
& \left(\begin{array}{c} u^c_{\alpha} \\ d^c_{\alpha} \end{array}\right)_R~,~
\left(\begin{array}{c} c^c_{\alpha} \\ s^c_{\alpha} \end{array}\right)_R~,~
\left(\begin{array}{c} t^c_{\alpha} \\ b^c_{\alpha} \end{array}\right)_R~,~
\left(\begin{array}{c} \nu_{e}^c \\ e^c \end{array}\right)_R~,~
\left(\begin{array}{c} \nu_{\mu}^c \\ \mu^c \end{array}\right)_R~,~
\left(\begin{array}{c} \nu_{\tau}^c \\ \tau^c \end{array}\right)_R~,~
\label{2.14}\\ 
& & \nonumber\\
& \begin{array}{c} (u^c_{\alpha})_L, \\ (d^c_{\alpha})_L, \end{array}~
\begin{array}{c} (c^c_{\alpha})_L, \\ (s^c_{\alpha})_L, \end{array}~
\begin{array}{c} (t^c_{\alpha})_L  \\ (b^c_{\alpha})_L, \end{array}~
(e^c)_L,~(\mu^c)_L,~(\tau^c)_L,
\label{2.15}
\eea
where $\alpha=1,2,3$ is the colour index. This exhausts all the 90
physical fermionic degrees of freedom. However, when ${\cal H}_F$ is
tensored with $L^2(S_M)$, the number of degrees of freedom becomes
redundant. In particular, an element $h_F \in {\cal H}_F$ can be
decomposed as
\be
h_F = h_L+h_R+h^c_L+h^c_R~~~,
\label{2.16}
\ee
where the four vectors in r.h.s. of (\ref{2.16}) belong to the
corresponding Hilbert spaces ${\cal H}_L$, ${\cal H}_R$, ${\cal
H}^c_L$ and ${\cal H}^c_R$, respectively. Furthermore, for each $x \in M$
a generic spinor $\psi$ can be decomposed as
\be
\psi(x) = \psi_L + \psi_R + \psi_R^c + \psi_L^c~~~.
\label{2.17}
\ee
Thus, by tensoring (\ref{2.17}) with (\ref{2.16}) we have 16 possible
combinations, namely four times the needed ones. In order to analyze
the physical meaning of these combinations it is useful to divide the
tensor product $\psi \otimes h_F$ in three parts 
\be
(\psi_L \otimes h_L + \psi_R \otimes h_R + \psi_R^c \otimes h_R^c + 
\psi_L^c \otimes h_L^c)~~~,\label{2.18}
\ee 
\be
(\psi_L \otimes h_R^c + \psi_R \otimes h_L^c + \psi_R^c \otimes h_L + 
\psi_L^c \otimes h_R)~~~,\label{2.19}
\ee
\be
(\psi_L+\psi_R^c)  \otimes (h_R + h_L^c) + (\psi_R+\psi_L^c)
\otimes (h_L+h^c_R)
~~~.\label{2.20}
\ee
The fermions in this last expression behave as the mirror fermions present
in lattice chiral gauge theories. In fact, if we consider for example the
term $\psi_L \otimes h_R$ of (\ref{2.20}) it corresponds to a left-handed
particle (as specified by $\psi_L$) which behaves under the gauge group as
a right-handed one (as specified by $h_R$). On the contrary, the
other two combinations (\ref{2.18}) and (\ref{2.19}) have the right
properties, though each of them independently is sufficient to describe
all the physical particles. As it will be clear in the next section, this
last redundancy is usually eliminated by identifying the degrees of
freedom of (\ref{2.18}) with the ones of (\ref{2.19}). Concerning the
unphysical part, it has to be projected out.

Let us denote with $P$ the projector on the physical subspace
corresponding to the combinations (\ref{2.18}) and (\ref{2.19}). Note that
this projector cannot be used from the very beginning. The above subspace,
in fact, would be no more a tensor product involving the space of Dirac
spinors, but rather Weyl spinors, and this would lead, as stated above, to
a trivial result. Furthermore, since $P$ does not commute with the
generalized Dirac operator and consequently with the gauge connection, the
form of the action would depend on the particular step of the construction
in which it has been used.

In literature, this problem seems to be ignored. What is implicitly done,
is to compute the trace of $\theta^2$ on the whole Hilbert space to get
the bosonic action, while the fermionic part is obtained with the {\em ad
hoc} prescription of retaining in the scalar product \eqn{2.6a} only the
physical state contribution.

This operation can be formally viewed as follows. First we note that the 
scalar product \eqn{2.6a} can be seen as a trace:
\be
{\cal S_F}=\mbox{\rm Tr} |\psi\rangle\langle(D+A+JAJ)\psi|~~~,
\ee
and we define the map $\Phi : {\cal H} \otimes \Omega^1 
\longrightarrow {\cal B(H)} $ 
\be
\Phi(\psi, A) = \theta^2(A) + |\psi\rangle\langle(D+A+JAJ)\psi|~~~,
\ee
where $\Omega^1$ is the space of 1-form connections.
The action is then the functional $\mbox{Tr} \circ \Phi|_{{\cal H}_{phys}
\otimes
\Omega^1}$, i.e. only the physical fermions are retained, while the trace
is still performed on the whole space ${\cal H}$.
This procedure is extremely ad hoc.

A possibility would then be to compute the traces on the physical 
Hilbert space. The action is then
\be
{\cal S}=\mbox{Tr} P\theta^2 + 
\mbox{Tr}|P\psi\rangle\langle(D+A+JAJ)P\psi|~~~.
\ee
Note that the only difference with respect to the usual action, is
in the presence of $P$ even for the bosonic term. Remarkably, this
difference for chiral theories, as we will see in the next section, will
cause the disappearance of some gauge physical degrees of freedom as well.

\section{Physical Hilbert space and the topological terms in chiral
gauge models}
In order to illustrate the general discussion of the previous section  
we consider first a simple model with gauge group $SU(2)_L
\otimes SU(2)_R$ and then the Standard Model, 
limiting ourselves to
the one quark family case to keep notations to a minimum.

\subsection{ $SU(2)_L \otimes SU(2)_R$ model}
In this case the algebra can be chosen as $\alg =
C^{\infty} (M, \complex) \otimes \alg_F$ where
$C^{\infty}(M,\complex)$ is the algebra of smooth complex valued
function on $M$ and $\alg_F = \quater_L \oplus \quater_R$, with
$\quater$ the algebra of quaternions. The Hilbert space is the tensor
product ${\cal H}$ = $L^2(S_M) \otimes {\cal H}_F$ and is the space of
spinor fields of the model. In particular we take ${\cal H}_F = {\cal
H}_L \oplus {\cal H}_R$ $ = \complex^2 \oplus \complex^2$
corresponding to two doublets $\xi_L$ and $\xi_R$ under,
respectively, the action of $\quater_L$ and $\quater_R$. Finally for
the generalized Dirac operator $D$ (\ref{2.3}), $D_F$ is chosen as the
mass matrix 
\begin{equation}
D_{F} = \left( \begin{array} {cc}
0 & {\cal M} \\
{\cal M}^{\dag} & 0 \end{array} \right)~~~.
\label{mass}
\ee
Having assigned the spectral triple $(\alg, {\cal H}, D)$ the model is
completely defined once a (faithful) representation of the algebra
on ${\cal H}$ is specified , which we choose as follows 
\be
\rho(q_L,q_R) \equiv
\left( \begin{array} {cc}
q_L & 0 \\
0 & q_R\\ \end{array} \right)~~~.
\label{rep1}
\ee
where $q_{L,R}$ are quaternions represented as $2 \times 2$ matrices.
Notice that in the general case one should also consider, as elements
of the Hilbert space, the charge conjugated states $\xi^c_{L,R}$,
related to the $\xi_{L,R}$ via the real structure $J$. This is only
necessary when more than one factor of the
chosen algebra is acting on each multiplet of the Hilbert space. This
is actually the case of the strong and electroweak Standard Model,
since, for example, left-handed quark doublets transform both under
the algebras $\quater_L$ and $M_3(\complex)$, whose unimodular elements
correspond to the groups $SU(2)_L$ and $SU(3)_c$. In our simple
example this is redundant,
since a simple representation of the algebra can be achieved and no
bivector potentials are required. We will come back to the more
general structure later, when we will consider the case of the
Standard Model. For the moment we are only interested in showing how
projecting out the unphysical degrees of freedom corresponds to the
natural appearance of topological terms in the classical action.

The gauge connection, computed according to (\ref{2.4}), takes the 
form\footnote{Concerning the Euclidean Dirac $\gamma$-matrices
we choose the hermitian representation $\gamma_{\mu}^{\dag}=
\gamma_{\mu}$. Moreover by definition $\gamma^{\mu \nu} \equiv
[\gamma^{\mu},\gamma^{\nu}]/2$.}:
\be
A(A_L,A_R,\phi) = \left( \begin{array} {cc}
 A\!\!\!/_L & \gamma_5 ( \phi - \phi_0) \\
\gamma_5 ( \phi^{\dagger} - \phi_0^{\dagger}) & A\!\!\!/_R 
\end{array} \right)~~~,
\label{1form}
\ee
where we have defined $A\!\!\!/_{L,R}  = \sum_i A\!\!\!/_{L,R}^i \sigma^i/2
\equiv \sum_i q^{'i}_{L,R}
\partial\!\!\!/ q^i_{L,R}$ and $\phi - \phi_0 \equiv
\sum_i q^{'i}_L ( {\cal M} q^i_R - q^i_L {\cal M} )$, where $\sigma^i$ are
the Pauli matrices, with the
condition $A^{\mu*}_{L,R}= - A^{\mu}_{L,R}$.
Note that, under a unimodular element $u$ of the algebra 
$A(A_L,A_R,\phi)$ transforms as
\be
A(A_L,A_R,\phi) \rightarrow u [ D,u] + u A(A_L,A_R,\phi) u^*~~~,
\ee
and therefore, using the representation for the algebra, it follows
that $A_{L,R}$ transform, as usual, as the adjoint representation of
the corresponding $SU(2)$ factor, and the Higgs field $\phi$ as a
doublet under both $SU(2)_L$ and $SU(2)_R$.
The corresponding curvature 2-form $\theta$, once the junk forms have
been subtracted out, reads
\be
\theta = \left( \begin{array} {cc}
\frac{1}{2} \gamma_{\mu \nu} F_L^{\mu \nu} + (\phi^{\dagger} \phi
- \phi^{\dagger}_0 \phi_0)
& - \gamma_5 {\cal D}\!\!\!/~ \phi \\
\gamma_5 ({\cal D}\!\!\!/~ \phi)^{\dagger} & 
\frac{1}{2} \gamma_{\mu \nu} F_R^{\mu \nu} + (\phi \phi^{\dagger}
- \phi_0 \phi^{\dagger}_0 ) \end{array} \right) ~~~,
\ee
where $F_{L,R}^{\mu \nu}$ are the usual gauge field tensors and
${\cal D}\!\!\!/~ \phi =(\partial\!\!\!/ + A\!\!\!/_L) \phi - \phi
A\!\!\!/_R$
is the covariant derivative of the Higgs field $\phi$.
Finally the bosonic action is calculated as $\mbox{Tr} \theta^2$, where the
trace is understood over the {\it internal} gauge degrees of freedom
and the external ones, related to the manifold $M$, which produces the
integration $\int d^4x$. The fermionic action is the scalar product
$\langle \psi, (D + A(A_L,A_R,\phi) )\psi \rangle$. Actually this last
contribution can be cast, as already mentioned, in the form of a trace of
an operator as well
\be
\langle \psi, (D + A(A_L,A_R,\phi) )\psi \rangle =
\mbox{Tr}[ \vert \psi \rangle \langle (D + A(A_L,A_R,\phi) \psi \vert]~~~.
\label{trsp}
\ee
We therefore obtain as classical action for this model
\bea
{\cal S} & = & \int d^4 x ~[ \frac{1}{4} \mbox{tr}F_L^{\mu \nu}
F^L_{ \mu \nu}
+ \frac{1}{4}\mbox{tr}F_R^{\mu \nu} F^R_{ \mu \nu} + ({\cal D}_{\mu} \phi)
( {\cal D}^{\mu} \phi)^{\dagger}
+ (\phi^{\dagger} \phi - \phi_0^{\dagger} \phi_0)^2 \nonumber \\
& + &
\overline{\Psi}_L (\partial\!\!\!/ + A\!\!\!/_L )
\Psi_L
+ \overline{\Psi}_R (\partial\!\!\!/ + A\!\!\!/_R )
\Psi_R
+ ( \overline{\Psi}_L ( {\cal M} + \phi ) \Psi_R + h. c.)]~~~,
\label{action}
\eea
where  $\Psi_{L,R} = \chi \otimes \xi_{L,R}$, $\chi$ a Dirac spinor
of $L^2(S_M)$ $\xi_i$ an element of ${\cal H}_F$, and tr denotes the
trace over the gauge internal indices.

As we have already noticed in the previous section the action
obtained contains a {\it doubling} problem, since both
right-handed and left-handed spinors $\chi$ of $L^2(S_M)$ couple to
the chiral gauge bosons.
The first procedure outlined there, which consists in simply neglecting in
the action the {\it wrong} states of the form $\chi_L \otimes
\xi_R$ and $\chi_R \otimes \xi_L$, leads to the customary result.

Let us now, instead, consider in more detail the second one, in which 
the trace in the action is restricted to the physical subspace via the
introduction of the projector operator $P$
\be
P = \frac{1 - \gamma_5}{2} \otimes {\cal P}_1 \oplus \frac{1 + \gamma_5}{2}
\otimes {\cal P}_2~~~,
\ee
where ${\cal P}_i$ is the projector onto
the component $\xi_i$ in the finite Hilbert space. In matrix form
\be
P = \frac{1}{2} \left( \begin{array} {cc}
1 - \gamma_5 & 0 \\
0 &  1 + \gamma_5 \\ \end{array} \right)~~~.
\ee
Note that $P$ commute with the curvature tensor $\theta$, as it can be
immediately checked by using the properties of gamma matrices. This means
that $\mbox{Tr}(P \theta P)^2$ $= \mbox{Tr} P \theta^2 P $ $= \mbox{Tr}
\theta^2 P $. After a straightforward computation
we have for the action of the model in this case
\bea
{\cal S} & = & \int d^4 x ~[ \frac{1}{4} \mbox{tr}[F_L^{\mu \nu}
F^L_{ \mu \nu} - F_L^{\mu \nu} \mbox{*}F_L^{\mu \nu}]
+ \frac{1}{4}\mbox{tr}[F_R^{\mu \nu} F^R_{ \mu \nu}
+ F_R^{\mu \nu} \mbox{*}F_R^{\mu \nu} ] \nonumber \\
& + & ({\cal D}_{\mu} \phi)
( {\cal D}^{\mu} \phi)^{\dagger}
+ (\phi^{\dagger} \phi - \phi_0^{\dagger} \phi_0)^2 \nonumber \\
& + &
\overline{\psi}_L (\partial\!\!\!/ + A\!\!\!/_L )
\psi_L
+ \overline{\psi}_R (\partial\!\!\!/ + A\!\!\!/_R )
\psi_R
+ ( \overline{\psi}_L ( {\cal M} + \phi ) \psi_R + h. c.)]~~~,
\label{actionp}
\eea
where *$F_{L,R}^{\mu \nu}=(1/2) \epsilon^{\mu \nu \rho \sigma}
F^{L,R}_{\rho \sigma}$ are the dual gauge tensor fields, and $\psi_{L,R} =
\chi_{L,R} \otimes \xi_{L,R}$ are the physical fermionic states. As a result
of the projection, all unphysical states in ${\cal S}$
disappear. On the other hand, in the gauge sector, only the antiself-dual
component of the gauge tensor field $F_L^{\mu \nu}$, 
satisfying (in the Euclidean space)
$F_L^{\mu \nu} = -\mbox{*}F_L^{\mu \nu}$
appears , while the self-dual has a vanishing kinetic term. Similarly
for $F_R^{\mu \nu}$ only the self-dual part
$F_R^{\mu \nu} = \mbox{*}F_R^{\mu \nu}$ contributes.
In other words the projection over states of definite chirality in the
fermionic sector leads to the result that a similar projection is made
onto the gauge fields, of which only one of the two independent component,
{\it left-moving} or {\it right-moving}, remains as a physical degree of
freedom, the other being projected out.

This result is of course unacceptable
from a physical point of view, since the action (\ref{actionp}) violates
$CPT$ symmetry. As we will see in the next section, in fact, similar
results hold in the case of the Standard Model, and no such dramatic
violation of $CPT$ symmetry are allowed by the huge phenomenology on
electroweak processes at low scale. It is worth noticing however that in the
case
of a purely vector or axial coupling of fermions to gauge fields,
with $SU(2)_V$ or $SU(2)_A$ gauge symmetry, namely if
one make the identification $F_L^{\mu \nu} = \pm F_R^{\mu \nu}$, the two
topological terms cancel each other and the corresponding action reduce
to the usual form. This is
the case, for example, for $SU(3)_c$ color interaction. For all chiral
gauge models, however, this procedure of projecting out unphysical
degrees of freedom in ${\cal H}$ at the level of curvature tensor $\theta$
gives a wrong result. 

\subsection{ The Standard Model }
For simplicity we will discuss the Connes-Lott version of the gauge theory
$SU(3)_c \otimes SU(2)_L \otimes U(1)_Y$ with only one quark family.
Actually the inclusion of leptons is quite crucial in order to obtain the
correct assignment of hypercharge quantum number
for fermions and Higgs bosons, by applying the
unimodularity condition. See for example \cite{Cordelia}.
Our discussion, however, is quite general and the results reported
can be easily generalized
when leptons and the correct number of fermion generations are considered.

The algebra in this case is chosen as $C^{\infty} (M, \complex)
\otimes \alg_F$, with
$\alg_F = M_3 \oplus \quater \oplus \complex$, with $M_3$
and $\quater$ the algebra of 3$\times$3 complex matrices and of
quaternions, respectively. The fermion Hilbert space is again the tensor
product ${\cal H} = L^2(S_M) \otimes {\cal H}_F$, where the finite factor
${\cal H}_F$ can be obtained from (\ref{2.7})-(\ref{2.11})
by only considering the first term in the direct sums and choosing $N=1$.
In particular it has as basis elements
the $SU(2)_L$ doublet $q_L^{\alpha}$, the $SU(2)_L$ singlets
$u_R^{\alpha}$ and $d_R^{\alpha}$, which we will collectively denote by
$q_R^{\alpha}$, and the corresponding $C$-conjugate states.
With $\alpha$ we denote the colour index.
Finally
the $D_F$ term in the Dirac operator (\ref{2.3}) is the fermion mass matrix
\be
D_F =
\left( \begin{array} {cccc}
0 & {\cal M} & 0 & 0 \\
{\cal M}^{\dagger}& 0 & 0 & 0 \\
0 & 0 & 0 & {\cal M}^* \\
0 & 0 & {\cal M}^T & 0 \\
\end{array} \right)~~~,
\label{quarkmass}
\ee
with
\be
{\cal M} = \left( \begin{array} {cc}
m_u \otimes \I_3 & 0 \\
0 & m_d \otimes \I_3 \end{array} \right)~~~.
\ee
The Connes-Lott representation of the algebra is the following
\be
\rho(c,q,B) \equiv
\left( \begin{array} {cccc}
q \otimes \I_3 & 0 & 0 & 0 \\
0 & B \otimes \I_3 & 0 & 0 \\
0 & 0 & \I_2 \otimes c & 0 \\
0 & 0 & 0 & \I_2 \otimes c \\
\end{array} \right)~~~,
\label{repsm}
\ee
with $c$, $q$ elements, respectively, of $M_3$, $\quater$ and
\be
B= \left( \begin{array} {cc}
b & 0 \\
0 & b^* \end{array} \right) ~~~ b \in \complex ~~~.
\ee
The calculation then goes along the same lines of the previous section.
In particular for the connection $A$ we get
\be
A(G,A_L,B,\phi) =
\left( \begin{array} {cccc}
A\!\!\!/_L \otimes \I_3 & \gamma_5 (\phi - \phi_0) \otimes \I_3
& 0 & 0 \\
\gamma_5 (\phi^{\dagger} - \phi^{\dagger}_0) \otimes \I_3& B\!\!\!/
\otimes \I_3 & 0 & 0 \\
0 & 0 & \I_2 \otimes G\!\!\!/ & 0 \\
0 & 0 & 0 & \I_2 \otimes G\!\!\!/ \\
\end{array} \right)~~~,
\ee
where as before $A\!\!\!/_L = \sum_i A\!\!\!/_L^i (\sigma_i/2)$ and
$G\!\!\!/ = \sum_{a=1}^8 G\!\!\!/^a (\lambda_a/2) + \I_3 G\!\!\!/^0$
with $\lambda_a$ the Gell-Mann $SU(3)$ matrices and
\be
B\!\!\!/ = \left( \begin{array} {cc}
\beta\!\!\!/ & 0 \\
0 & \beta\!\!\!/^* \end{array} \right) ~~~ \beta_{\mu} \in \complex~~~.
\ee
From the conditions
$A_L^{\mu*} = - A_L^{\mu}$, $G^{\mu*} = - G^{\mu}$ and
$B^{\mu*} = - B^{\mu}$ in particular it follows  $\beta_{\mu} = 
- \beta^*_{\mu}$

For the curvature tensor, with all junk forms subtracted out, one has instead
\be
\theta = \left( \begin{array} {cccc}
\frac{1}{2} \gamma_{\mu \nu} F_L^{\mu \nu} + (\phi^{\dagger} \phi
- \phi^{\dagger}_0 \phi_0)
& - \gamma_5 {\cal D}\!\!\!/~ \phi & 0 & 0 \\
\gamma_5 ( {\cal D}\!\!\!/~ \phi)^{\dagger}
& \frac{1}{2} \gamma_{\mu \nu} B^{\mu \nu} + (\phi
\phi^{\dagger} - \phi_0 \phi_0^{\dagger}) & 0 & 0 \\
0 & 0 & \frac{1}{2} \gamma_{\mu \nu} G^{\mu \nu} & 0 \\
0 & 0 & 0 & \frac{1}{2} \gamma_{\mu \nu} G^{\mu \nu} \\
\end{array} \right)~~~,
\ee
with ${\cal D}\!\!\!/~ \phi = (\partial\!\!\!/ + A\!\!\!/_L) \phi - \phi
B\!\!\!\!/$.
The unimodularity condition $\mbox{Tr} (A + J A J)=0$ removes the $U(1)$
factor corresponding to $G^0_{\mu}$ and the action, obtained by tracing
over the entire Hilbert space, reads
\bea
{\cal S} & = & \int d^4 x ~[ \frac{1}{4} \mbox{tr}F_L^{\mu \nu}
F^L_{ \mu \nu}
+ \frac{1}{4} B^{\mu \nu} B_{ \mu \nu}
+ \frac{1}{4}\mbox{tr}G^{\mu \nu} G_{ \mu \nu}
+ ({\cal D}_{\mu} \phi)
( {\cal D}^{\mu} \phi)^{\dagger}
+ (\phi^{\dagger} \phi - \phi_0^{\dagger} \phi_0)^2 \nonumber \\
& + &
\overline{Q}_L (\partial\!\!\!/ + A\!\!\!/_L + G\!\!\!/)
Q_L +
\overline{Q}^c_R (\partial\!\!\!/ + A\!\!\!/_L + G\!\!\!/)
Q^c_R 
 +  \overline{Q}_R (\partial\!\!\!/ + B\!\!\!/ + G\!\!\!/ ) Q_R
\nonumber \\
& + & \overline{Q}^c_L (\partial\!\!\!/ - B\!\!\!/ + G\!\!\!/ ) Q^c_L
 + 
( \overline{Q}_L ( {\cal M} + \phi ) Q_R +
\overline{Q}^c_R ( {\cal M}^* + \phi ) Q^c_L + h. c.) ]~~~,
\label{actionsm}
\eea
where  $Q_{L,R} = \chi \otimes q_{L,R}^{\alpha}$, with
$\chi$ a Dirac spinor
of $L^2(S_M)$. The redundancy of the degrees of freedom due to the presence
of $C$-conjugate states can be eliminated by identifying the corresponding
states, $Q_{L,R} \equiv Q^c_{R,L}$, namely making the quotient with respect
to the equivalence relation given by the real structure $J$. On the other
hand, as in the case discussed in the previous section,
in the action appear mirror unphysical
states with a chirality {\it mismatch}, like $\chi_L \otimes q_R$ or
$\chi_R \otimes q_L$.

If we now restrict the trace of the operator $\theta^2 +
\vert \psi \rangle \langle (D + A(G,A_L,B,\phi) \psi \vert$
via the introduction of the projector $P$
\be
P = \frac{1}{2} \left( \begin{array} {cccc}
1 - \gamma_5 & 0 & 0 & 0 \\
0 & 1 + \gamma_5 & 0 & 0 \\
0 & 0 & 1 + \gamma_5 & 0 \\
0 & 0 & 0 & 1 - \gamma_5 \end{array} \right)
\ee
we get
\bea
{\cal S} & = & 
\int d^4 x ~[ \frac{1}{4} \mbox{tr}[F_L^{\mu \nu}
F^L_{ \mu \nu} - F_L^{\mu \nu} \mbox{*} F^L_{\mu \nu}]
+ \frac{1}{4} [B^{\mu \nu} B_{ \mu \nu} + B^{\mu \nu} \mbox{*} B_{ \mu \nu}]
+ \frac{1}{4}\mbox{tr} G^{\mu \nu} G_{ \mu \nu} \nonumber \\
& + & ({\cal D}_{\mu} \phi)
( {\cal D}^{\mu} \phi)^{\dagger}
+ (\phi^{\dagger} \phi - \phi_0^{\dagger} \phi_0)^2 \nonumber \\
& + &
\overline{q}_L (\partial\!\!\!/ + A\!\!\!/_L + G\!\!\!/)
q_L +
\overline{q}^c_R (\partial\!\!\!/ + A\!\!\!/_L + G\!\!\!/)
q^c_R 
 +  \overline{q}_R (\partial\!\!\!/ + B\!\!\!/ + G\!\!\!/ ) q_R
\nonumber \\
& + & \overline{q}^c_L (\partial\!\!\!/ - B\!\!\!/ + G\!\!\!/ ) q^c_L
+ ( \overline{q}_L ( {\cal M} + \phi ) q_R +
\overline{q}^c_R ( {\cal M}^* + \phi ) q^c_L + h. c.) ]~~~,
\label{actionsmp}
\eea
where, to simplify notation, we have denoted with $q_{L,R}$ the
physical states $\chi_{L,R} \otimes q_{L,R}^{\alpha}$.
Hence the topological terms appear for both the $SU(2)_L$
and the $U(1)$
factor, while the $SU(3)_c$ fields, due to its vector coupling to
quarks, contribute to ${\cal S}$ with the usual
$\mbox{tr} G^{\mu \nu} G_{ \mu \nu}$ term only. In particular
the self-dual component of $F_L$ receives no kinetic contribution and
is {\it projected out} by the introduction of $P$.
In the fermion sector, instead, all unphysical states are absent.
As already pointed out this approach, leading to (\ref{actionsmp}), or
(\ref{actionp}), implies $CPT$ symmetry violation and is at variance
with low-energy phenomenology.

\section{The Spectral Action}
Recently Chamseddine and Connes \cite{Connesgrav,ChamsConnes,MarsSpectr}
have proposed another form for the bosonic action, which includes gravity
as well, while the fermionic action remains the same.

The idea behind the spectral action is that while the topology is
encoded by the algebra, all other information (metric in first
instance) are encoded by the generalized covariant Dirac operator
$D_A=D+A+JAJ$, where now $D_A$ also contains the
spin connection terms. Moreover, the operator $D$ can
be characterized completely by its spectrum. This leads Connes and
Chamseddine to consider as bosonic action 
\be
{\cal S_B}=\mbox{Tr}\left[\chi\left( - {D_A^2\over m_0^2}\right)\right]~~~,
\label{4.1}
\ee
where $\chi$ is a suitable cutoff function.
The quantity $m_0$ is a cutoff with dimensions (in natural
units) of a mass which indicates at which scale the theory under
consideration effectively shows its noncommutative geometric
nature. The action basically sums up
the eigenvalues of $D_A$ which are smaller than $m_0$. The trace can be
evaluated with heath kernel techniques \cite{Gilkey}. We now consider
what happens in the model discussed in sections 3.1.
To further simplify notations we make the additional simplification of
only considering
the gauge and gravitational contribution. The other contributions can be 
added without altering the results. 
This is accomplished by choosing
a Dirac operator with vanishing fermion mass terms
\be
D_{A} = \left(\begin{array}{cc} D_L & 0 \\ 0 & D_R \end{array}
\right)~~~,
\label{4.2}
\ee
where 
\bea
D_L &\equiv& \nabla\!\!\!\!/ \otimes \I_2 - { i \over 2} g_L 
A\!\!\!/_L~~~,\label{4.3}\\
D_R &\equiv& \nabla\!\!\!\!/ \otimes \I_2 - { i \over 2} g_R 
A\!\!\!/_R~~~.\label{4.4}
\eea
Furthermore, by $\nabla_{\mu}$ we
denote the covariant derivative corresponding to the metric
connection only.

According to the considerations developed in the previous sections we
substitute the expression (\ref{4.1}) for the action, with a similar
expression in which the trace is performed over the physical states
only. This can be done by using the previously defined
projector $P$
\be
P = \left( \begin{array}{cc} P_{L} & 0 \\
0 & P_{R} \end{array} \right) = { 1 \over 2}
\left( \begin{array}{cc} 1 - \gamma_{5} & 0 \\
0 & 1 + \gamma_{5} \end{array} \right)
~~~.\label{4.2a}
\ee
Hence in terms of (\ref{4.2a}) we get
\bea
{\cal S_B} = \mbox{Tr}\left[\chi\left(-{D_A^2\over m_0^2}\right)P\right]
& = & \mbox{Tr}\left[\chi\left(-{D_A^2 P\over m_0^2}\right)\right]
\nonumber \\
 & = & \mbox{Tr}\left[\chi\left(-{D_L^2 P_{L}\over m_0^2}\right)\right] +
\mbox{Tr}\left[\chi\left(-{D_R^2 P_R \over m_0^2}\right)\right]
~~~,
\label{4.2b}
\eea
where to get the r.h.s. of (\ref{4.2b}) we have used the property the
$P$ commutes with $D^2$.
Note that the minus sign in (\ref{4.2b})
is due to our choice of hermitian $\gamma$-matrices.

The trace (\ref{4.2b}) is defined by using the heat-kernel expansion
\cite{Gilkey}
\be
\mbox{Tr}\left[\chi\left(-{D^2 P\over m_0^2}\right)\right] \simeq
\sum_{n \geq 0} f_{n} \left[a_{n}\left(-{D_L^2 P_L\over m_0^2}\right)
+a_{n}\left(-{D_R^2 P_{R}\over m_0^2}\right)\right]
~~~,
\label{4.2c}
\ee
where the coefficients $f_{n}$ are given by
\bea
f_{0} &=& \int_{0}^{\infty} \chi(u)~ u~ du~~~,~~~~~~~~~f_{2}
= \int_{0}^{\infty} \chi(u)~ du~~~,\nonumber\\
f_{2(n+2)}&=&(-1)^n~ \chi^{(n)}(0)~~~,~~~~\mbox{with}~~~~n\geq 0~~~,
\label{4.2d}
\eea
and 
\be
a_{n}\left(-{D_{L,R}^2 P_{L,R}\over m_0^2}\right) = \int_{M} \sqrt{g}~
a_{n}\left(x,-{D_{L,R}^2 P_{L,R}\over m_0^2}\right)~d^4x ~~~.
\label{4.2e}
\ee
Note that $a_{n}$ vanish for odd $n$.

From definitions (\ref{4.3}), (\ref{4.4}) and (\ref{4.2b})
one can compute the positive definite operator
\bea
- D_{L,R}^2 P_{L,R} &=& \left\{-\Box_{1/2} \otimes \I_2 - {1 \over 8} 
R_{\mu \nu \rho \sigma} \gamma^{\mu \nu} \gamma^{\rho \sigma} \otimes \I_2 +
{ i \over 2} g_{L,R} \left( D_{\mu}^{L,R} A^{\mu}_{L,R} \right) \right.
\nonumber\\
&+& \left.
{ i \over 4} g_{L,R} \gamma^{\mu\nu} F_{\mu \nu}^{L,R} + i g_{L,R} 
\I_{4} \otimes A_{L,R}^{\mu} \nabla_{\mu}\right\} P_{L,R}~~~,
\label{4.5}
\eea
where $D^{L,R}_{\mu} A^{L,R}_{\nu} = \nabla_{\mu} A^{L,R}_{\nu}
- i (g_{L,R}/2) [A^{L,R}_{\mu},A^{L,R}_{\nu}]$
is the complete covariant derivative and $ F_{\mu \nu}^{L,R} 
\equiv D^{L,R}_{\mu} A_{\nu}^{L,R} - D^{L,R}_{\nu} A_{\mu}^{L,R}$.
Moreover,  in the previous equation $\Box_{1/2} \equiv \nabla^{\mu} 
\nabla_{\mu} = g^{\mu \nu} (\partial_{\mu}
+ \omega_{\mu})(\partial_{\nu} + \omega_{\nu}) - \Gamma^{\mu}
(\partial_{\mu} +\omega_{\mu})$, where $\omega_{\mu}$ denotes the
spin connection, $\Gamma^{\mu} \equiv g^{\rho \sigma} 
\Gamma^{\mu}_{\rho \sigma}$, and finally, we have chosen the 
representation for the Riemann tensor according to which $R_{1212}=1/r^2$
on the 2d sphere of radius $r$. From these definitions  we can recast
Eq. (\ref{4.5}) as
\be
- D_{L,R}^2 P_{L,R}= - \left( g^{\mu \nu} P_{L,R}\otimes \I_2
\partial_{\mu} \partial_{\nu}
+C^{\mu}_{L,R} \partial_{\mu} + B_{L,R}\right)~~~,
\label{4.6}
\ee
where
\bea
C^{\mu}_{L,R} & = & \left[ \left( 2 \omega^{\mu} - \Gamma^{\mu}
\right) \otimes \I_{2} - i g_{L} \I_{4} \otimes A_{L,R}^{\mu} \right]
P_{L,R}
~~~,\label{4.7}\\
B_{L,R} & = & \left[\left( \partial_{\mu} \omega^{\mu} + \omega_{\mu}
\omega^{\mu} - { R \over 4}~\I_4 - \Gamma^{\mu} \omega_{\mu}\right)
\otimes \I_2 - i g_{L,R} \omega_{\mu} A_{L,R}^{\mu}\right.
\nonumber\\
&-& \left.{ i \over 2} g_{L,R} \I_4 \otimes 
\left( D_{\mu}^{L,R} A_{L,R}^{\mu}\right) 
- {i \over 4} g_{L} \gamma^{\mu \nu} F^{L,R}_{\mu \nu} \right]
P_{L,R}~~~.
\label{4.8}
\eea
In order to apply the formalism developed in Ref. \cite{Gilkey}
to compute the Seeley-deWitt coefficients, it is convenient to
introduce the following quantities 
\be
\xi^{\mu}_{L,R} \equiv  {1 \over 2} \left(C^{\mu}_{L,R} +
\Gamma^{\mu} P_{L,R}\right)
= \left(\omega_{\mu} \otimes \I_{2} - {i \over 2} g_{L,R} \I_{4}
A^{\mu}_{L,R} \right) P_{L,R}~~~,
\label{4.9}
\ee
\bea
E_{L,R} & \equiv & B_{L,R} - 
\left(\partial_{\mu} \xi^{\mu}_{L,R} + \xi^{\mu}_{L,R}\xi_{\mu}^{L,R}
- \Gamma^{\mu} \xi_{\mu}^{L,R} \right)\nonumber\\
&=& - \left[ {R \over 4} ~\I_4 \otimes \I_2 + { i \over 4} g_{L,R}
\gamma^{\mu \nu} F_{\mu \nu}^{L,R} \right] P_{L,R}~~~,
\label{4.10}
\eea
and 
\be
\Omega_{\mu \nu}^{L,R} = \left[{ 1 \over 4} 
R_{\mu \nu \rho \sigma} \gamma^{\rho
\sigma} \otimes \I_{2} - { i \over 2} g_{L,R} \I_{4} \otimes  
F^{L,R}_{\mu \nu} \right] P_{L,R}~~~.
\label{4.11}
\ee
Thus by following \cite{Gilkey} we get
\be
a_{0}\left(x,-{D_{L,R}^2 P_{L,R}\over m_0^2}\right) =
{1 \over 16 \pi^2} \mbox{Tr}\left(P_{L,R} \otimes \I_2\right)=
{1 \over 4 \pi^2}~~~,\label{4.12}
\ee
\be
a_{2}\left(x,-{D_{L,R}^2 P_{L,R}\over m_0^2}\right)
=  {1 \over 16 \pi^2} \mbox{Tr}\left(E_{L,R} + {R \over 6}~
P_{L,R} \otimes \I_2 \right)~~~,\label{4.13}
\ee
\bea
a_{4}\left(x,-{D_{L,R}^2 P_{L,R}\over m_0^2}\right)
= { 1 \over 16 \pi^2} { 1 \over 360}
\left[ 60 \Box E + 60 R E + 180 E^2
+ 30 \Omega_{\mu \nu} \Omega^{\mu \nu} \right.
\nonumber\\
+ \left.
\left(12 \Box R + 5 R^2  - 2 R_{\mu \nu} R^{\mu \nu}
+ 2 R_{\mu \nu \rho \sigma} R^{\mu \nu \rho \sigma} \right) P_{L,R}
\otimes \I_2 \right],
\label{4.14}
\eea
By substituting in Eq.s (\ref{4.12}-(\ref{4.14}) the expressions
(\ref{4.10}) and (\ref{4.11}) we get
\be
a_{0}\left(x,-{D_{L}^2 P_{L}\over m_0^2}\right) +
a_{0}\left(x,-{D_{R}^2 P_{R}\over m_0^2}\right) = 
{ 1 \over 2 \pi^2}~~~,
\label{4.15}
\ee
\be
a_{2}\left(x,-{D_{L}^2 P_{L}\over m_0^2}\right) +
a_{2}\left(x,-{D_{R}^2 P_{R}\over m_0^2}\right) = 
- { 1 \over 24 \pi^2} R~~~,
\label{4.16}
\ee
and
\bea
a_{4}\left(x,-{D_{L}^2 P_{L}\over m_0^2}\right) &+&
a_{4}\left(x,-{D_{R}^2 P_{R}\over m_0^2}\right) =  
{ 1 \over 16 \pi^2} {1 \over 180} \left\{ - 12 \Box R + 5 R^2 - 8 
R_{\mu \nu} R^{\mu \nu} \right. \nonumber \\
& - &   7 R_{\mu \nu \rho \sigma} R^{\mu \nu \rho \sigma}
+ 15 g_{L}^2 \mbox{Tr}\left(F_{\mu \nu}^L F^{\mu \nu}_{L} \right)
+ 15 g_{R}^2 \mbox{Tr}\left(F_{\mu \nu}^R F^{\mu \nu}_{R} \right) \nonumber
\\
& - & \left. {45 \over 8} g_{L}^2 \mbox{Tr}\left(F_{\mu \nu}^L 
\mbox{*} F^{\mu \nu}_{L} \right)
+{45 \over 8} g_{R}^2 \mbox{Tr}\left(F_{\mu \nu}^R 
\mbox{*} F^{\mu \nu}_{R} \right)\right\}~~~.
\label{4.17}
\eea
We can therefore conclude that also in the case of the spectral action
the effect of considering traces over the physical Hilbert space
is that of adding the topological term to the action. In this case, however,
unlike what happens in the (CL) case,
the two components of the gauge fields, the self-dual and antiself-dual,
are both present but their kinetic term are weighted by different factors.
Again this represents a violation of $CPT$ symmetry of the model.

\section{Conclusions}
In this paper we have analysed the structure of the Hilbert space ${\cal H}$
adopted in the noncommutative geometry models of gauge theories.
We have pointed out that ${\cal H}$ contains unphysical fermionic degrees 
of freedom. Some of these are harmless and can be removed with a quotient.
The others, behaving as mirror particles, can be eliminated only via a
projection. What seems to us unsatisfactory is that the projection
operator does not fit into the geometrical construction. In fact since it
does not commute with the generalized Dirac operator, it leads to
different results depending on the step in the construction
at which it is applied. Furthermore none of these correspond to a
physically acceptable model. The only way to get, for example, the
correct result for the standard model, is through the ad hoc prescription
of neglecting the unphysical fermionic degrees of freedom in the action.

How serious is this problem for the noncommutative geometry
approach to the standard model? One can take two extreme views. On one
side one can consider the fact that some extra terms in the lagrangian
appear to be irrelevant and simply ignore them without worrying too
much. On the other hand one can consider that noncommutative geometry fails
in its attempt to at least reproduce the action of a gauge
theory. Obviously both these positions can be easily 
criticised . Noncommutative
geometry remains a programme which is able to give very fruitful
results in particle physics and gravity, but probably,
in order to obtain the Standard Model in a fully
geometrical way, some modification of the spectral triple, or of some
other crucial ingredient of the theory, is needed.

\ \\
{\bf Acknowledgments}\\
\\
We would like to thank G.~Bimonte, G.~Esposito and G.~Landi
for useful discussions.
F. Lizzi and G. Miele would like to thank N.~Mavromatos for
hospitality in the
Physics Department (Theoretical Physics) of the University of Oxford.

% A macro to raise things. Used in math and journal macros.
\def\up#1{\leavevmode \raise.16ex\hbox{#1}}
%journal references
\newcommand{\npb}[3]{{\sl Nucl. Phys. }{\bf B#1} \up(19#2\up) #3}
\newcommand{\plb}[3]{{\sl Phys. Lett. }{\bf #1B} \up(19#2\up) #3}
\newcommand{\revmp}[3]{{\sl Rev. Mod. Phys. }{\bf #1} \up(19#2\up) #3}
\newcommand{\sovj}[3]{{\sl Sov. J. Nucl. Phys. }{\bf #1} \up(19#2\up) #3}
\newcommand{\jetp}[3]{{\sl Sov. Phys. JETP }{\bf #1} \up(19#2\up) #3}
\newcommand{\rmp}[3]{{\sl Rev. Mod. Phys. }{\bf #1} \up(19#2\up) #3}
\newcommand{\prd}[3]{{\sl Phys. Rev. }{\bf D#1} \up(19#2\up) #3}
\newcommand{\ijmpa}[3]{{\sl Int. J. Mod. Phys. }{\bf A#1} \up(19#2\up) #3}
\newcommand{\prl}[3]{{\sl Phys. Rev. Lett. }{\bf #1} \up(19#2\up) #3}
\newcommand{\physrep}[3]{{\sl Phys. Rep. }{\bf #1} \up(19#2\up) #3}
\newcommand{\jgp}[3]{{\sl J. Geom. Phys.}{\bf #1} \up(19#2\up) #3}
\newcommand{\journal}[4]{{\sl #1 }{\bf #2} \up(19#3\up) #4}

\end{document}